\documentstyle[12pt]{article}
\def\r#1{(\ref{#1})}
\def\Lms{\Lambda_{\overline{MS}}} 
\def\als{\alpha_s}
\newcommand{\nn}{\nonumber}
\newcommand{\mt}{m_{t}}
\newcommand{\mtb}{\overline{m}_{t}}

\newcommand{\mc}{m_{c}}
\newcommand{\ms}{m_{s}}
\newcommand{\md}{m_{d}}

\newcommand{\bea}{\begin{eqnarray}}
\newcommand{\eea}{\end{eqnarray}}
\newcommand{\be}{\begin{equation}}
\newcommand{\ee}{\end{equation}}
\newcommand{\bi}{\begin{itemize}}
\newcommand{\ei}{\end{itemize}}
\newcommand{\ord}{{\cal O}}

\newcommand{\epp}{\varepsilon^\prime/\varepsilon}
\newcommand{\ep}{\varepsilon_K}
\def\Im{\mathop{\mbox{Im}}}
\def\Re{\mathop{\mbox{Re}}}

\newcommand{\klpn}{K_L \to \pi^0 \nu \bar \nu}
\newcommand{\kppn}{K^+ \to \pi^+ \nu \bar \nu}
\newcommand{\kmm}{K_L \to \mu^+ \mu^-}
\newcommand{\kpe}{K_L \to \pi^0 e^+ e^-}

\def\npb#1#2#3{    {\it Nucl. Phys. }{\bf B #1} (19#2) #3}
\def\plb#1#2#3{    {\it Phys. Lett. }{\bf B #1} (19#2) #3}
\def\prd#1#2#3{    {\it Phys. Rev. }{\bf D #1} (19#2) #3}

\def\prl#1#2#3{    {\it Phys. Rev. Lett. }{\bf #1} (19#2) #3}

\def\epjc#1#2#3{   {\it Eur. Phys. J. }{\bf C #1} (19#2) #3}

\begin{document}
\thispagestyle{empty}
\begin{flushright}
  TUM-HEP-334/98 \\
 November 1998
\end{flushright}
\vskip 1truecm 
\centerline{\Large\bf Upper Bounds on $K \to \pi \nu \bar \nu$ and
  $\kpe$}
\centerline{\Large\bf from $\epp$ and $K_L \to \mu^+
  \mu^-$\footnote[1]{\noindent Supported by the German
    Bundesministerium f{\"u}r Bildung und Forschung under contract 06 TM
    874 and by the DFG project Li 519/2-2.}}  
\vskip1truecm
\centerline{\large\bf Andrzej J. Buras and Luca Silvestrini} \bigskip
\centerline{\sl Technische Universit{\"a}t M{\"u}nchen, Physik Department}
\centerline{\sl D-85748 Garching, Germany} \vskip1truecm
\centerline{\bf Abstract} 
We analyze rare kaon decays in models in which the dominant new effect
is an enhanced $\bar s d Z$ vertex $Z_{ds}$. We point out that in
spite of large theoretical uncertainties the CP-violating ratio $\epp$
provides at present the strongest constraint on $\Im Z_{ds}$. Assuming
$0 \le \epp \le 2 \cdot 10^{-3}$ and Standard Model values for the CKM
parameters we obtain the bounds ${\rm BR}(\klpn)\le 2.4 \cdot
10^{-10}$ and ${\rm BR}(\kpe) \le 3.6 \cdot 10^{-11}$ (which are
substantially stronger than the bounds found recently by Colangelo and
Isidori, using $\ep$ instead of $\epp$). We illustrate how these
bounds can be improved with the help of the forthcoming data on
$\epp$. Using the bound on $\Re Z_{ds}$ from $\kmm$ we find ${\rm
  BR}(\kppn) \le 2.3 \cdot 10^{-10}$. In this context we derive an
analytic upper bound on ${\rm BR}(\kppn)$ as a function of ${\rm
  BR}(\klpn)$ and the short distance contribution to ${\rm BR}(\kmm)$.
We also discuss new physics scenarios in which in addition to an
enhanced $\bar s d Z$ vertex also neutral meson mixing receives
important new contributions. In this case larger values of the
branching ratios in question cannot be excluded.
\vfill 
\newpage

\section{Introduction}
\label{sec:intro}

Flavour-Changing Neutral Current (FCNC) processes provide a powerful
tool for the tests of the Standard Model and the physics beyond it. Of
particular interest are the rare kaon decays $\klpn$, $\kppn$ and $\kpe$
which are governed by $Z$-penguin diagrams. Within the Standard Model
the branching ratios for these decays are known including
next-to-leading order corrections \cite{nlo1,nlo2}. Updating the analysis
in \cite{bratios} we find
\begin{eqnarray}
  \label{eq:brklpn}
  {\rm BR}(\klpn)&=&(3.1 \pm 1.3) \cdot 10^{-11}\,, \\
  \label{eq:brkpnn}
  {\rm BR}(\kppn)&=&(8.2 \pm 3.2) \cdot 10^{-11}\,, \\
  {\rm BR}(\kpe)_{\rm dir}&=&(4.9 \pm 2.1) \cdot 10^{-12}\,, 
  \label{eq:brkpe}
\end{eqnarray}
where the errors come dominantly from the uncertainties in the CKM
parameters. The corresponding theoretical uncertainties in
\r{eq:brklpn}-\r{eq:brkpe} amount to at most a few percents.  The
reduction of ${\rm BR}(\kppn)$ and the increase of ${\rm BR}(\klpn)$
and ${\rm BR}(\kpe)$ with respect to the values given in
\cite{bratios} originate primarily from the improved experimental
lower bound on the $B^0_s$--$\bar B^0_s$ mixing and the slight
increase of $|V_{ub}|$ \cite{RUDO}.  The branching ratio in
eq.~\r{eq:brkpe} represents the so-called direct CP-violating
contribution to $\kpe$. The remaining two contributions to this decay,
the CP-conserving one and the indirect CP-violating one, are plagued by
theoretical uncertainties \cite{KL}.  They are expected to be $\ord
(10^{-12})$ but generally smaller than ${\rm BR}(\kpe)_{\rm dir}$.
This implies that within the Standard Model ${\rm BR}(\kpe)$ is
expected to be at most $10^{-11}$.

Experimentally we have \cite{kplus}
\begin{equation}
  \label{eq:brkpnnexp}
  {\rm BR}(\kppn) = (4.2 \,^{+9.7}_{-3.5})\cdot 10^{-10}
\end{equation}
and the bounds \cite{klong,ke} 
\begin{equation}
  \label{eq:brklexp}
  {\rm BR}(\klpn) < 1.6 \cdot 10^{-6}\,, 
  \qquad {\rm BR}(\kpe) < 4.3 \cdot 10^{-9}.
\end{equation}
Moreover from \r{eq:brkpnnexp} and isospin symmetry one has
\cite{isospin} ${\rm BR}(\klpn) < 6.1 \cdot 10^{-9}$. The data on these
three branching ratios should improve considerably in the coming
years.

In this context a very interesting claim has been recently made by
Colangelo and Isidori \cite{CI}, who analyzing rare kaon decays in
supersymmetric theories pointed out a possible large enhancement of
the effective $\bar s d Z$ vertex leading to branching ratios as high
as 
\begin{eqnarray}
  {\rm BR}(\klpn) &=& 4 \cdot 10^{-9}\,, \nn \\
  {\rm BR}(\kppn) &=& 1 \cdot 10^{-9}\,, \nn \\
  {\rm BR}(\kpe)_{\rm dir} &=& 6 \cdot 10^{-10}\,. 
  \label{eq:isidori}
\end{eqnarray}

This amounts to an enhancement of ${\rm BR}(\kppn)$ by one order of
magnitude and of ${\rm BR}(\klpn)$ and ${\rm BR}(\kpe)$ by two orders
of magnitude relative to the Standard Model expectations. According to
\cite{CI} such enhancements are still compatible with data for other
FCNC processes such as $\kmm$, the $K_L$--$K_S$ mass difference
$\Delta M_K$ and the indirect CP violation in $K_L \to \pi \pi$
represented by the parameter $\ep$. Not surprisingly these results
brought a lot of excitement among experimentalists.

In this paper we would like to point out that in models in which the
dominant new effect is an enhanced $\bar s d Z$ vertex such large
enhancements of ${\rm BR}(\klpn)$ and ${\rm BR}(\kpe)$ are already
excluded by the existing data on the CP-violating ratio $\epp$ in
spite of the large theoretical uncertainties. Similarly the large
enhancement of ${\rm BR}(\kppn)$ can be excluded by the data on $\epp$
and in particular by the present information on the short distance
contribution to $\kmm$. The latter can be bounded by analysing the
data on ${\rm BR}(\kmm)$ in conjunction with improved estimates of
long distance dispersive contributions \cite{dambrosio,pich}.

Our main point is as follows. Similarly to the rare decays in question
also $\epp$ depends sensitively on the size of $Z$-penguin
contributions and generally on the size of the effective $\bar s d Z$
vertex. In the Standard Model $Z$-penguins dominate the so-called
electroweak penguin contributions to $\epp$ which enter this ratio
with the opposite sign to QCD penguins and suppress considerably
$\epp$ for large $m_t$ \cite{flynn}.  If the new physics beyond the
Standard Model does not modify significantly the QCD penguin
contributions but enhances considerably the $Z$-penguin contribution,
then the strong cancellation between QCD and electroweak penguin
contributions to $\epp$ present in the Standard Model does not take
place. Consequently large positive or negative values of $\epp$
(dependently on the sign of new contributions) can be found. While
precise predictions for $\epp$ cannot be made due to substantial
hadronic uncertainties, the very fact that the cancellation in
question is removed makes the estimates more reliable.  Assuming that
the usual analysis of the unitarity triangle involving neutral meson
mixing is only weakly modified by new physics contributions we find
that the values in \r{eq:isidori} simultaneously imply $\vert \epp
\vert > 5 \times 10^{-3}$. This is by many standard deviations higher
than the present average $(1.5 \pm 0.8) \times 10^{-3}$ extracted
\cite{PDG} from the experiments at CERN \cite{cern} and Fermilab
\cite{fermi}. As we will demonstrate below, if $0 \le (\epp)_{\rm exp}
\le 2 \cdot 10^{-3}$ then
\begin{equation}
  \label{eq:bounded}
  {\rm BR}(\klpn) \le 2.4 \cdot 10^{-10} \qquad {\rm and} \qquad
  {\rm BR}(\kpe)_{\rm dir} \le 3.6 \cdot 10^{-11}\,,
\end{equation}
which is a factor of 20 smaller than what anticipated in \cite{CI}.
The new round of $\epp$ measurements could improve these bounds
considerably. In this context one should note that $\epp$ is linear in
the imaginary part of the $\bar s d Z$ vertex, whereas ${\rm
  BR}(\klpn)$ and ${\rm BR}(\kpe)_{\rm dir}$ are quadratically
dependent on this vertex. Thus a modest reduction of the possible
range for $(\epp)_{\rm exp}$ would result in a substantial improvement
of the bounds in \r{eq:bounded}.

The bound on ${\rm BR}(\kppn)$ is governed dominantly by the bound on
${\rm BR}(\kmm)_{\rm SD}$ as we will explicitly demonstrate below.
Reanalyzing this bound we find that the upper
bound on the real part of the $\bar s d Z$ vertex has been
overestimated in \cite{CI} by roughly a factor of two. Combining
this finding with the analysis of $\epp$ we find
\begin{equation}
  \label{eq:boundedp}
  {\rm BR}(\kppn) \le 2.3 \cdot 10^{-10}\,
\end{equation}
which is roughly a factor of four lower than given in \cite{CI}. We
would like to stress that further improvements on $(\epp)_{\rm exp}$
will have only a minor impact on \r{eq:boundedp}. On the other hand
improved data on ${\rm BR}(\kmm)$ and in particular further
improvements in the theoretical analysis of the long distance
dispersive contribution to $\kmm$ could have an important impact on
the bound in \r{eq:boundedp}.

We also analyze a scenario of new physics in which there are
substantial new contributions to neutral meson mixing so that
the usual analysis of the unitarity triangle does not apply. Dependent
on the sign of the new contributions the bounds on rare decays become
stronger or weaker. The largest branching ratios are found when 
neutral meson mixing is dominated by new physics contributions that
are twice as large and have opposite sign with respect to the Standard
Model. This possibility is quite remote. However, if this situation
could be realized in some exotic model, then the branching ratios in
question could be as high as
\begin{eqnarray}
  {\rm BR}(\klpn) &=& 1.4 \cdot 10^{-9}\,, \nn \\
  {\rm BR}(\kppn) &=& 5.3 \cdot 10^{-10}\,, \nn \\
  {\rm BR}(\kpe)_{\rm dir} &=& 2.2 \cdot 10^{-10}\,, 
  \label{scenb}
\end{eqnarray}
which are still lower than given in \r{eq:isidori}. On the other hand
if the CKM matrix is assumed to be real we find the maximal values to
be
\begin{eqnarray}
  {\rm BR}(\klpn) &=& 3.7 \cdot 10^{-10}\,, \nn \\
  {\rm BR}(\kppn) &=& 2.9 \cdot 10^{-10}\,, \nn \\
  {\rm BR}(\kpe)_{\rm dir} &=& 6.6 \cdot 10^{-11}\,.
  \label{scenc}
\end{eqnarray}

The paper is organized as follows. In Section \ref{sec:strategy} we
summarize our strategy and present the basic formulae. In Section
\ref{sec:bounds} we derive the bounds on the effective $\bar s d Z$
vertex using $\kmm$ and $\epp$.  In this context we derive also an
analytic upper bound on ${\rm BR}(\kppn)$ as a function of ${\rm
  BR}(\klpn)$ and the short distance contribution to ${\rm BR}(\kmm)$.
Subsequently we present implications of these bounds for $\kppn$,
$\klpn$ and $\kpe$. In Section \ref{sec:summary} we summarize briefly
our main findings.

\section{Basic Formulae}
\label{sec:strategy}

We are interested in the one-loop flavour-changing effective coupling
of the $Z$-boson to down-type quarks, in the limit of vanishing
external masses and momenta. The corresponding effective Lagrangian
can be generally written as
\begin{equation}
  \label{eq:Wds}
  {\cal L}^Z_{\rm FC} = \frac{G_F}{\sqrt{2}} \frac{e}{2 \pi^2} M_Z^2
  \frac{\cos \Theta_W}{\sin \Theta_W} Z_{ds} \bar s \gamma_\mu
  (1-\gamma_5) d~ Z^\mu\,+\, {\rm h.c.}
\end{equation}
where $Z_{ds}$ is a complex coupling. In the Standard Model one has
\begin{equation}
  \label{eq:Wsm}
  Z_{ds}^{\rm SM} = \lambda_t C_0(x_t)\,, 
\qquad x_t=\frac{\mtb^2}{M_W^2}\,,
\end{equation}
where $\lambda_t=V_{ts}^* V_{td}$ with $V_{ij}$ being the CKM matrix
elements. $C_0(x_t)$ is a real function which for the central value
of the top quark mass, 
$\mtb(\mt)=166$ GeV, equals
$0.79$. Its explicit expression can be found in \cite{bratios}.

From the standard analysis of the unitarity triangle, we find
\begin{equation}
  \label{eq:smlt}
  \Im \lambda_t = (1.38 \pm 0.33) \cdot 10^{-4}\,, \qquad \Re
  \lambda_t = -(3.2 \pm 0.9) \cdot 10^{-4}\,, 
\end{equation}
and consequently 
\begin{equation}
  \label{eq:smW}
  \Im Z_{ds}^{\rm SM} = (1.09 \pm 0.26) \cdot 10^{-4}\,, \qquad \Re
  Z_{ds}^{\rm SM} = -(2.54 \pm 0.71) \cdot 10^{-4}\,, 
\end{equation}
where the error in $\mt$ has been neglected.

For completeness we give the relation of $Z_{ds}$ to $U_{ds}$ defined
by 
\begin{equation}
  \label{eq:Uds}
  {\cal L}^Z_{\rm tree} = \frac{g_2}{4 \cos \Theta_W} U_{ds} \bar s
  \gamma_\mu (1-\gamma_5) d~ Z^\mu\,+\, {\rm h.c.}
\end{equation}
and used in the extensions of the Standard Model in which tree level
flavour-changing $Z^0$ couplings appear \cite{silvn,silverman,ds2}. 
One has
\begin{equation}
  \label{eq:UdsWds}
  U_{ds} = \frac{\sqrt{2} G_F M_W^2}{\pi^2} Z_{ds} = 1.08 \cdot
  10^{-2} Z_{ds}\,.
\end{equation}
Our effective coupling $Z_{ds}$ includes both Standard Model and new
physics contributions. This definition is convenient for our analysis,
since it allows us to take automatically into account possible
interference effects between the Standard Model and new physics.
Notice however that the coupling $W_{ds}$ used in ref.~\cite{CI}, and
in particular the bounds derived in \cite{CI} from $\kmm$, $\kppn$ and
$\ep$, do not include the Standard Model contribution. Care must be
therefore taken in comparing our bounds on $Z_{ds}$ with the bounds on
$W_{ds}$ of ref.~\cite{CI}.

In obtaining the formulae listed below we have used the following
strategy. In the case of $\klpn$, $\kppn$, $\kpe$, $(\kmm)_{\rm SD}$
and $\epp$ we have taken the Standard Model expressions and replaced
there $\lambda_t C_0$ by $Z_{ds}$. The remaining contributions
resulting from box diagrams, gluon-penguin and photon-penguin diagrams
have been evaluated for $\mtb(\mt)=166$ GeV. That is we assume that
new physics will have at most a minor impact on the latter
contributions. This is for example the case in supersymmetric
extensions of the Standard Model. As pointed out in ref.~\cite{CI}, in
general supersymmetric models there is the interesting possibility of
a substantial enhancement of the effective $\bar s d Z$ vertex with
respect to its Standard Model value, via a double helicity-flipping
flavour-changing squark mass insertion. On the other hand, the effect
of all other supersymmetric contributions to $K \to \pi \nu \bar \nu$
decays via penguin and box diagrams can at most be of the order of
magnitude of the Standard Model contribution. Details can be found in
ref.~\cite{BRS}. Furthermore, it has been shown that in general
supersymmetric models the gluonic penguin contributions to $\epp$ are
small due to the constraints coming from $\Delta M_K$ and $\ep$, and
to the negative interference between $\Delta S=1$ box and penguin
diagrams \cite{GMS}. The same applies to photonic penguins.  We can
therefore conclude that the only place where supersymmetry can produce
order-of-magnitude enhancements in the above-mentioned processes is an
enhanced $\bar s d Z$ vertex via the Colangelo-Isidori mechanism.

We also assume that no new operators in addition to those present in
the Standard Model contribute. In this case the replacement $\lambda_t
C_0 \to Z_{ds}$ is justified without the modification of QCD
renormalization group effects evaluated at NLO level for scales below
$\ord (\mt)$.

It should be remarked that $B_0$ and $C_0$ depend on the gauge
parameter in the W-propagator. This gauge dependence cancels when both
are taken into account. In using $Z_{ds}$ instead of $C_0$ we assume
that the latter function is hidden in $Z_{ds}$ so that the gauge
dependence in question is absent in the final formulae for branching
ratios. However, it is present in $Z_{ds}$. The values for $Z_{ds}$
quoted in this paper correspond then to `t Hooft-Feynman gauge. In any
case, this gauge dependence is rather weak as it originates in terms
which are non-leading in $\mt$.  As box diagrams contributing to rare
decays receive only small contributions from physics beyond the
Standard Model we expect that the gauge dependence of new
contributions to $Z_{ds}$ is also very weak.

In deriving the formula for $\epp$ we use the NLO analytic formula for
this ratio \cite{anal1} which has been updated in \cite{bratios}. Since
$\epp$ depends visibly on $\Lms^{(4)}$ we identify those coefficients in
the formula in question which carry the dominant $\Lms^{(4)}$
dependence.  We evaluate the remaining ones for $\Lms^{(4)}=325$ MeV,
corresponding to $\als^{\overline {MS}}(M_Z)=0.118$. This leaves us
first with two coefficients $r_0^{(6)}$ and $r_Z^{(8)}$ which in the range
$0.113 \le \als^{\overline {MS}}(M_Z) \le 0.123$ are related within
$10 \%$ by $r_0^{(6)} \simeq 1.1 \vert r_Z^{(8)}\vert$. We will use this
relation in what follows. The resulting formula for $\epp$ reproduces
in the limit $Z_{ds} \to Z_{ds}^{\rm SM}$ the Standard Model results
within $10 \%$ accuracy, which is clearly sufficient for our purposes.

Our basic formulae read then as follows:
\begin{eqnarray}
  {\rm BR}(\kppn) &=& 1.55 \cdot 10^{-4} \Biggl[ \biggl( \Im Z_{ds}-4 B_0
  \Im \lambda_t\biggr)^2 \nn \\
  &&\qquad + \biggl(\Re Z_{ds} + \Delta_c - 4 B_0 \Re
  \lambda_t\biggr)^2\Biggr]\,, 
  \label{eq:fkppn}
\end{eqnarray}
where 
\begin{equation}
  \label{eq:deltac}
  \Delta_c = - (2.11 \pm 0.30) \cdot 10^{-4}
\end{equation}
represents the internal charm contribution \cite{nlo1} and
$B_0=-0.182$ is the box diagram function evaluated at $\mtb(\mt)=166$
GeV.  New physics contributions are expected to arise at a scale $\ge
M_W$ and are therefore included in $Z_{ds}$, while the charm
contribution, generated at the scale $m_c$, can be safely evaluated in
the Standard Model.  Next
\begin{eqnarray}
  \label{eq:fklpn}
  {\rm BR}(\klpn) &=& 6.78 \cdot 10^{-4} \Bigl[ \Im Z_{ds} - 4 B_0 \Im
  \lambda_t \Bigr]^2\,, \\
  \label{eq:fkpe}
  {\rm BR}(\kpe) &=& 1.19 \cdot 10^{-4} \biggl[ \Bigl( \Im Z_{ds} - B_0 \Im
  \lambda_t\Bigr)^2\\ 
  &&\qquad + \Bigl(\Im \lambda_t + 0.08 \Im Z_{ds}\Bigr)^2\biggr]\,, \nn \\
  \label{eq:fkmm}
  {\rm BR}(\kmm)_{\rm SD} &=& 6.32 \cdot 10^{-3} \Bigl[\Re Z_{ds} - B_0 \Re
  \lambda_t + \bar \Delta_c \Bigr]^2 \,,
\end{eqnarray}
where
\begin{equation}
  \label{eq:deltacbar}
  \bar \Delta_c = - (6.54 \pm 0.60) \cdot 10^{-5}\,.
\end{equation}
represents the charm contribution \cite{nlo1}.

Using \r{eq:fkppn}, \r{eq:fklpn} and \r{eq:fkmm} we derive the 
following useful formula 
\begin{eqnarray}
  {\rm BR}(\kppn) &=& 1.55 \cdot 10^{-4} 
\Biggl[ \pm 3.97\sqrt{\kappa}\cdot 10^{-4}-3 B_0 \Re\lambda_t+
 \hat\Delta_c\Biggr]^2 \nn \\
  &&\qquad + 0.229\cdot  {\rm BR}(\klpn),
   \label{eq:fkppf}
\end{eqnarray}
where
\begin{equation}
\label{eq:deltahat}
 \hat\Delta=\Delta_c-\bar\Delta_c=- (1.46 \pm 0.30) \cdot 10^{-4}\,
\end{equation}
and $\kappa$ is defined through
\begin{equation}
\label{kappa}
  {\rm BR}(\kmm)_{\rm SD}=  \kappa \cdot 10^{-9}.
\end{equation}
In evaluating $\hat\Delta_c$ we have included the correlation between
$\Delta_c$ and $\bar\Delta_c$ due to their simultaneous dependence
on $\Lms^{(4)}$ and $\mc$ \cite{nlo1}.
Next we have
\begin{eqnarray}
  \label{eq:fdmk}
  \Delta M_K &=& \left( \Delta M_K \right)^{\rm SM} + 
  8.2 \cdot 10^{-12} \eta_{QCD}\hat B_K \biggl[ \left(\Re Z_{ds} \right)^2 -
  \left(\Im Z_{ds} \right)^2 \biggr]\,, \\
  \label{eq:fep}
  \ep &=& \ep^{\rm SM} - 1.65 \cdot 10^{3} \left(\Re Z_{ds} \right)
  \left(\Im Z_{ds} \right)\eta_{QCD} \hat B_K e^{i \pi/4}\,.
\end{eqnarray}
Here $(\Delta M_K)^{\rm SM}$ and $\ep^{\rm SM}$ are the Standard Model
box contributions and the second terms in \r{eq:fdmk} and \r{eq:fep}
stand for $Z^0$ contributions.  $\eta_{QCD}$ is the QCD factor and
$\hat B_K$ is a hadronic parameter.  One has $\eta_{QCD} \hat B_K
\approx 0.5$.

In the limit $\eta_{QCD} \hat B_K =1$ the $Z^0$ contributions given
here are larger by a factor of two relatively to the ones presented in
\cite{CI,silvn}. They agree on the other hand with those presented in
\cite{ds2}. For our purposes, however, this factor is irrelevant since
$\Delta M_K$ and $\ep$ provide much weaker constraints on the coupling
$Z_{ds}$ than $\kmm$ and $\epp$.

Finally we decompose $\epp$ as follows:
\begin{equation}
  \label{eq:eppdec}
  \frac{\varepsilon^\prime}{\varepsilon} =
  \left(\frac{\varepsilon^\prime}{\varepsilon}   \right)_Z
  +\left(\frac{\varepsilon^\prime}{\varepsilon}   \right)_{\rm Rest} 
\end{equation}
and proceeding as explained above we find
\begin{eqnarray}
  \label{eq:eppz}
  \left(\frac{\varepsilon^\prime}{\varepsilon}   \right)_Z &=& \Im
  Z_{ds} \Bigl[ 1.2 - R_s \vert r_Z^{(8)}\vert B_8 ^{(3/2)}\Bigr]\,, \\
  \left(\frac{\varepsilon^\prime}{\varepsilon}   \right)_{\rm Rest}
  &=& \Im \lambda_t \biggl[-2.3 + R_s \Bigl[1.1 \vert r_Z^{(8)}\vert
  B_6^{(1/2)} + (1.0 + 0.12 \vert r_Z^{(8)}\vert )
  B_8^{(3/2)}\Bigr]\biggr]\,.\nn 
\end{eqnarray}
Here
\begin{equation}
  \label{eq:rs}
  R_s = \left[ \frac{158 {\rm MeV}}{\ms(\mc) + \md(\mc)} \right]^2\,.
\end{equation}
$B_6^{(1/2)}$ and $B_8^{(3/2)}$ are non-perturbative parameters
describing the hadronic matrix elements of the dominant QCD-penguin
and electroweak penguin operators respectively. Finally $\vert
r_Z^{(8)}\vert$ is a calculable renormalization scheme independent
 parameter in the analytic formula for
$\epp$ in \cite{bratios} which increases with $\als^{\overline {MS}}(M_Z)$
and in the range $0.113 \le \als^{\overline {MS}}(M_Z) \le 0.123$ takes
the values 
\begin{equation}
  \label{eq:rz8}
  6.5 \le  \vert r_Z^{(8)}\vert \le 8.5\,.
\end{equation}
For $R_s$ we will use the range
\begin{equation}
  \label{eq:rrs}
  1 \le R_s \le 2\,,
\end{equation}
which is compatible with the most recent lattice and QCD sum rules
calculations as reviewed in \cite{gupta}. We consider the ranges
in \r{eq:rz8} and \r{eq:rrs} as conservative. Similarly we will use
\begin{equation}
  \label{eq:bpars}
  0.8 \le B_6^{(1/2)} \le 1.2\,, \qquad 0.6 \le B_8^{(3/2)} \le 1.0\,,
\end{equation}
which is compatible with the recent lattice and large N calculations 
as reviewed in \cite{bratios,gupta}. 
Our treatment of $\Im \lambda_t$ and $\Re\lambda_t$ will be explained
below.

\section{Bounds from $\kmm$ and $\epp$}
\label{sec:bounds}
\subsection{Generalities}

In deriving the bounds on $\Re Z_{ds}$ and $\Im Z_{ds}$ from $\kmm$
and $\epp$ we have to investigate whether $\lambda_t$ extracted from
the standard analysis of the unitarity triangle and given in
\r{eq:smlt} is significantly modified by new physics contributions.
After the improved lower bound on $B_s^0$--$\bar B^0_s$ mixing
\cite{RUDO}, the determination of $\lambda_t$ in the standard analysis
is governed by the values of $\vert V_{ub}\vert$, $\vert V_{cb}\vert$,
$\ep$ and the ratio of $B_d^0$--$\bar B^0_d$ and $B_s^0$--$\bar B^0_s$
mixings, represented by $(\Delta M)_d/(\Delta M)_s$. The values of
$\vert V_{ub} \vert$ and $\vert V_{cb} \vert$ are expected to be
unaffected by new physics contributions. On the other hand, new
physics can easily affect neutral meson mixings. We can therefore
distinguish between three different scenarios:
\begin{itemize}
\item[A)] New physics only manifests itself through an enhanced $\bar
  s d Z$ vertex. In this case, the bounds on $\Re Z_{ds}$ from $\kmm$
  and on $\Im Z_{ds}$ from $\epp$ imply that the $Z^0$ contribution to
  $\Delta M_K$ is fully negligible and the corresponding contribution
  to $\ep$ is below 15\% of the experimental value. Consequently, the
  standard analysis of the unitarity triangle is only marginally
  affected by new physics and we can use for $\lambda_t$ the values
  given in eq.~\r{eq:smlt}.
\item[B)] New physics affects the $\Delta F=2$ amplitudes through some
  mechanism other than $Z^0$-exchange (for example, this is what
  happens in general SUSY models such as the one considered by
  Colangelo and Isidori). In this case the standard analysis of the
  unitarity triangle is invalid. In order to be as general as
  possible, in this scenario we only assume unitarity of the CKM
  matrix. In this case
  \begin{equation}
    \label{eq:imltgen}
    \Im \lambda_t = \vert V_{ub} \vert \vert V_{cb} \vert \sin
    \delta\,, 
  \end{equation}
  with $0 \le \delta \le 2 \pi$ and \cite{BLO}
   \begin{equation}
   \label{eq: relt}
   \left (\lambda-\frac{\lambda^3}{2}\right)|V_{cb}|_{\rm min}^2 
   (1-R^{\rm max}_b)
   \le -\Re\lambda_t \le
   \left(\lambda-\frac{\lambda^3}{2}\right)|V_{cb}|_{\rm max}^2 
   (1+R^{\rm max}_b)
  \end{equation}
   where $\lambda=0.22$ is the Wolfenstein parameter and
  \begin{equation}
\label{rb}
   R_b= \left(1-\frac{\lambda^2}{2}\right)\frac{1}{\lambda}
        \left|\frac{V_{ub}}{V_{cb}}\right|.
   \end{equation}
\item[C)] Scenario B with $\Im\lambda_t=0$. Here CP-violation
 originates entirely in new physics contributions.
  \end{itemize}

In the following two subsections we discuss the constraints on 
$Z_{ds}$ from $\kmm$ and $\epp$ in scenario A, and their
implications for rare $K$ decays. Subsequently we will discuss
briefly scenarios B and C.

\subsection{$\kmm$ and the Bounds on $\kppn$ and $\Re Z_{ds}$}

The $\kmm$ branching ratio can be decomposed generally as follows:
\begin{equation}
  \label{eq:deckmm}
  {\rm BR}(\kmm)=\vert \Re A \vert^2 + \vert \Im A \vert^2\,,
\end{equation}
where $\Re A$ denotes the dispersive contribution and $\Im A$ the
absorptive one. The latter contribution can be determined in a model
independent way from the $K_L \to \gamma \gamma$ branching ratio. The
resulting $\vert \Im A \vert^2$ is very close to the experimental
branching ratio ${\rm BR}(\kmm)=(7.2 \pm 0.5) \cdot 10^{-9}$ 
\cite{mpmm} so that $\vert
\Re A \vert^2$ is substantially smaller and extracted to be \cite{mpmm}
\begin{equation}
  \label{eq:reaexp}
  \vert \Re A_{\rm exp} \vert^2 < 5.6 \cdot 10^{-10} \qquad {\rm 
  (90\% \, \,C.L.).}
\end{equation}
Now $\Re A$ can be decomposed as 
\begin{equation}
  \label{eq:decomp}
  \Re A = \Re A_{\rm LD} + \Re A_{\rm SD}\,,  
\end{equation}
with 
\begin{equation}
  \label{eq:reasdbr}
  \vert \Re A_{\rm SD} \vert^2 \equiv {\rm BR}(\kmm)_{\rm SD}
\end{equation}
representing the short-distance contribution given in \r{eq:fkmm}. An
improved estimate of the long-distance contribution $\Re A_{\rm LD}$
has been recently presented by D'Ambrosio, Isidori and Portol{\'e}s
\cite{dambrosio} who find
\begin{equation}
  \label{eq:reald}
  \vert \Re A_{LD} \vert < 2.9 \cdot 10^{-5} \qquad {\rm (90\% \,\,C.L.).} 
\end{equation}
The highest possible value for $BR(\kmm)_{\rm SD}$ is found if $\Re
A_{\rm SD}$ and $\Re A_{LD}$ have opposite sign. The bounds
\r{eq:reaexp} and \r{eq:reald} give then
\begin{equation}
  \label{eq:sdlimitbr}
  {\rm BR}(\kmm)_{\rm SD} < 2.8 \cdot 10^{-9}.
\end{equation}
This result is very close to the one presented very recently
by Gomez Dumm and Pich \cite{pich}. The bound in \r{eq:sdlimitbr}
should be compared with the short distance contribution within
the Standard Model for which we find
\begin{equation}
\label{kmusm}
 {\rm BR}(\kmm)^{\rm SM}_{\rm SD}=(9.4\pm4.1)\cdot 10^{-10}.
\end{equation}

Due to the presence of the charm contribution, the constraint on $\Re
Z_{ds}$ that one can extract from \r{eq:sdlimitbr} depends on the sign
of $\Re Z_{ds}$. From \r{eq:fkmm} we get
\begin{equation}
  \label{eq:rewdslim}
   -5.6 \cdot 10^{-4} \le \Re Z_{ds} \le 8.1 \cdot 10^{-4}.
\end{equation}
The upper bound is obtained using $\Re\lambda_t=-4.1\cdot 10^{-4}$
and the lower one using $\Re\lambda_t=-2.3\cdot 10^{-4}$. Note
that the term $\bar\Delta_c$ as well as $B_0\Re\lambda_t$ play only
a minor role in obtaining \r{eq:rewdslim}. Setting them to zero
we would find $|\Re Z_{ds}| \le 6.7 \cdot 10^{-4}$.

The bound in \r{eq:rewdslim} is compatible with the one of 
Silverman \cite{silverman}
who finds $\vert \Re Z_{ds} \vert \le 8.3 \cdot 10^{-4}$. To
compare with ref.~\cite{CI} one has to add the Standard Model
contribution, since the bound $\vert \Re W_{ds}\vert \le 2.2 \cdot
10^{-3}$ in eq.~(3.10) of ref.~\cite{CI} only refers to the new
physics contribution. We would therefore get, neglecting the charm
contribution, 
\begin{equation}
  \label{eq:confisi}
  \Re W_{ds} < 6.7 \cdot 10^{-4} + \vert \Re Z_{ds}^{\rm SM} \vert \le
  10^{-3},  
\end{equation}
which is roughly a factor of two stronger than the bound in
ref.~\cite{CI}.\footnote{This difference can be traced to a factor of
  two missing in eq.~(9) of ref.~\cite{dambrosio}, as confirmed by the
  authors.}

Having the upper bound on ${\rm BR}(\kmm)_{\rm SD}$ we can use the formula 
\r{eq:fkppf} to derive an upper bound on ${\rm BR}(\kppn)$. This bound is
obtained by choosing the negative sign in \r{eq:fkppf}, which corresponds
to $\Re Z_{ds}< 0.$ Setting $\Re\lambda_t=-4.1 \cdot 10^{-4}$,
$\hat\Delta_c=-1.76\cdot 10^{-4}$ we obtain
\begin{equation}
{\rm BR}(\kppn)\le 0.229\cdot {\rm BR}(\klpn)
+2.44\cdot 10^{-11}[\sqrt{\kappa}+1.01]^2.
\end{equation}
With $\kappa=2.8$ (see \r{eq:sdlimitbr}) this reduces to

\begin{equation}
\label{bound}
{\rm BR}(\kppn)\le 0.229\cdot {\rm BR}(\klpn)+1.76\cdot 10^{-10}.
\end{equation}

This formula allows then to find the upper bound on ${\rm BR}(\kppn)$
once the upper bound on ${\rm BR}(\klpn)$ is known. As we will now
demonstrate, the latter bound can be obtained from $\epp$.

\subsection{Constraints from $\epp$ in scenario A}
\label{sec:caseA}

In this scenario all the effects of new physics are encoded in the
effective coupling $Z_{ds}$. As we shall verify explicitly, in this
case the new physics effects in $\Delta M_K$ and $\ep$ can be
neglected and we can use the values for $\lambda_t$ given in
eq.~\r{eq:smlt}. 

The form of the bound for $\Im Z_{ds}$ from $\epp$ depends on the sign
of $\Im Z_{ds}$. We consider first the case $\Im Z_{ds} < 0$. Then
$(\epp)_Z$ in \r{eq:eppz} is positive and adds up to $ (\epp)_{\rm
  Rest}$ which is also positive. The upper bound on $-\Im Z_{ds}$
reads then 
\begin{equation}
  \label{eq:imwds1}
  -\Im Z_{ds} \le \frac{\left(\frac{\varepsilon^\prime}{\varepsilon}
  \right)^{\rm exp}_{\rm max} - \left(\frac{\varepsilon^\prime}{\varepsilon}
  \right)_{\rm Rest}}{\left(R_s \vert r_Z^{(8)} \vert B_8^{(3/2)} -
  1.20\right) } 
\end{equation}
The most conservative bound is found by setting $\Im \lambda_t = 1.05
\cdot 10^{-4}$, $R_s=1$, $B_6^{(1/2)}=0.8$, $B_8^{(3/2)}=0.6$ and
$\vert r_Z^{(8)} \vert = 6.5$. This gives 
\begin{equation}
  \label{eq:imwds2}
  -\Im Z_{ds} \le \frac{\left(\frac{\varepsilon^\prime}{\varepsilon}
  \right)^{\rm exp}_{\rm max} - 4.71 \cdot 10^{-4}}{2.70}. 
\end{equation}
Setting for instance $(\epp)^{\rm exp}_{\rm max} = 2 \cdot 10^{-3}$ we
find 
\begin{equation}
  \label{eq:imwds3}
  -\Im Z_{ds} \le 5.66 \cdot 10^{-4}. 
\end{equation}
If $\Im Z_{ds} > 0$ the bound \r{eq:imwds1} changes to 
\begin{equation}
  \label{eq:imwds4}
  \Im Z_{ds} \le \frac{\left(\frac{\varepsilon^\prime}{\varepsilon}
  \right)_{\rm Rest} - \left(\frac{\varepsilon^\prime}{\varepsilon}
  \right)^{\rm exp}_{\rm min}}{\left(R_s \vert r_Z^{(8)} \vert B_8^{(3/2)} -
  1.20\right) } 
\end{equation}
as for large $\Im Z_{ds}$ $\epp$ becomes negative and what counts is
the minimal value of $(\epp)^{\rm exp}$.
The most conservative bound is obtained by setting
$B_6^{(1/2)}=1.2$, $B_8^{(3/2)}=0.6$ and $\Im \lambda_t = 1.71 \cdot
10^{-4}$, and varying $r_Z^{(8)}$ and $R_s$ in the ranges \r{eq:rz8} and
\r{eq:rrs}. It turns out that the dependence of the bound on these two
parameters is rather weak. For $(\epp)^{\rm exp}_{\rm min}=0$ we find 
\begin{equation}
  \label{eq:imwds5}
  \Im Z_{ds} \le 4.65 \cdot 10^{-4},
\end{equation}
which is close to the bound in \r{eq:imwds3}.

Clearly the bounds on $\Im Z_{ds}$ depend on the experimental values
of $(\epp)^{\rm exp}_{\rm max}$ and $(\epp)^{\rm exp}_{\rm min}$. We
illustrate this in table \ref{tab:eppex}. In parentheses we show
the bounds one would obtain for $B_8^{(3/2)}=1.0$. Clearly in this
case the upper bounds are much stronger. The dependence of the
bounds on $B_6^{(1/2)}$ is weaker than on $B_8^{(3/2)}$. 
Needless to say a reduction of the uncertainties in the values of 
$B_6^{(1/2)}$ and $B_8^{(3/2)}$ is clearly desirable.

\begin{table}
  \begin{center}
    \begin{tabular}{||c|c|c|c||}
      \hline \hline
      $(\epp)^{\rm exp}_{\rm max}$ & $2 \cdot 10^{-3}$ & $1.5 \cdot
      10^{-3}$ & $1 \cdot 10^{-3}$ \\ \hline 
      $\left(-\Im Z_{ds}\right)_{\rm max}[10^{-4}]$ 
      & $5.7~(2.7) $ & $3.8~(1.8)$ & $2.0~(0.9)$ \\ \hline \hline
      $(\epp)^{\rm exp}_{\rm min}$ & $0$ & $5 \cdot
      10^{-4}$ & $1 \cdot 10^{-3}$ \\ \hline
      $\left(\Im Z_{ds}\right)_{\rm max}[10^{-4}]$ 
      & $4.7~(2.7)$ & $3.7~(2.3)$ & $3.2~(2.0)$ \\ \hline \hline
    \end{tabular}
    \caption{Constraints on $\Im Z_{ds}$ from $\epp$ in scenario A
     for different
      values of $(\epp)^{\rm exp}_{\rm max}$ and $(\epp)^{\rm
        exp}_{\rm min}$ with $B_8^{(3/2)}=0.6~(1.0).$} 
    \label{tab:eppex}
  \end{center}
\end{table}

Using now the constraints on $\Im Z_{ds}$ in table \ref{tab:eppex} we
can compute the upper bounds on the rare decays $\klpn$ and $\kpe$.
Similarly using \r{bound} we can calculate the upper bound on $\kppn$.
These bounds are given in tables \ref{tab:rarepos} and
\ref{tab:rareneg} for positive and negative $\Im Z_{ds}$ respectively.
These tables illustrate how the improvement in the allowed range for
$\epp$, to be expected in the near future, and the increase of the
minimal value of $B_8^{(3/2)}$ could improve the bounds in question.
The most conservative bounds in these tables are listed in
\r{eq:bounded} and \r{eq:boundedp}.

\begin{table}[htbp]
\begin{center}
  \begin{tabular}{||c|c|c|c||}
  \hline \hline
  $(\epp)^{\rm exp}_{\rm min}$ &  $0$ & $5 \cdot
  10^{-4}$ & $1 \cdot 10^{-3}$ \\ \hline 
  ${\rm BR}(\klpn)[10^{-10}]$ & $2.4~(1.0)$ &$1.7~(0.9)$ & 
   $1.3~(0.7)$\\ \hline 
  ${\rm BR}(\kpe)[10^{-11}]$ & $3.5~(1.5)$ &$ 2.4~(1.2)$ &
  $1.9~(1.0)$ \\ \hline 
  ${\rm BR}(\kppn)[10^{-10}]$ & $2.3~(2.0)$ & $2.1~(1.9)$ &
  $2.1~(1.9)$ \\ \hline \hline
  \end{tabular}
  \caption{Upper bounds for the rare decays $\klpn$, $\kppn$ and
    $\kpe$, obtained in scenario A by imposing the constraints in
    eq.~\r{bound} and in table \ref{tab:eppex}, in the case $\Im
    Z_{ds}>0$ with $B_8^{(3/2)}=0.6~(1.0).$}
  \label{tab:rarepos}
\end{center}
\end{table}

\begin{table}[htbp]
  \begin{center}
    \begin{tabular}{||c|c|c|c||}
      \hline \hline
  $(\epp)^{\rm exp}_{\rm max}$ &  $2 \cdot 10^{-3}$ & $1.5 \cdot 10^{-3}$
  & $1 \cdot 10^{-3}$ \\ \hline 
  ${\rm BR}(\klpn)[10^{-10}]$ & $1.6~(0.3)$ & $0.6~(0.1)$&
  $0.1~(0.0)$ \\ \hline 
  ${\rm BR}(\kpe)[10^{-11}]$ & $3.6~(0.9)$ & $1.6~(0.4)$ &
  $0.5~(0.2)$ \\ \hline 
  ${\rm BR}(\kppn)[10^{-10}]$ & $2.1~(1.8)$ & $1.9~(1.8)$ &
  $1.8~(1.8)$ \\ \hline \hline
    \end{tabular}
    \caption{Upper bounds for the rare decays $\klpn$, $\kppn$ and
    $\kpe$, obtained in scenario A by imposing the constraints in
    eq.~\r{bound} and in table \ref{tab:eppex}, in the case $\Im
    Z_{ds}<0$ with $B_8^{(3/2)}=0.6~(1.0).$}
  \label{tab:rareneg}
  \end{center}
\end{table}

The contributions due to $Z_{ds}$ to $\Delta M_K$ are always
negligible, while in $\ep$ the effects of $Z_{ds}$ can reach $15 \%$,
and can therefore be neglected in the standard analysis of the
unitarity triangle.

\subsection{Constraints from $\epp$ in scenario B}
\label{sec:caseB}

We now want to discuss the most general case in which new physics
contributions to neutral meson mixing can be so large as to completely
invalidate the standard analysis of the unitarity triangle. In this
case, we just impose unitarity of the CKM matrix and let 
$\lambda_t$ vary in the range
\begin{equation}
  \label{eq:ltgen}
  -2 \cdot 10^{-4} \le \Im\lambda_t \le 2 \cdot 10^{-4}\,.
\end{equation}
\begin{equation}
  \label{eq:rtgen}
  1.54 \cdot 10^{-4} \le -\Re\lambda_t \le 5.85 \cdot 10^{-4}\,.
\end{equation}
These ranges are obtained from \r{eq:imltgen}-\r{rb} 
using $V_{cb}=0.040\pm0.003$ and
$|V_{ub}/V_{cb}|=0.091\pm 0.016$ \cite{RUDO}.

For positive $\Im Z_{ds}$ the upper bound on $\Im Z_{ds}$ is as in
\r{eq:imwds4} the only modification being $(\Im\lambda_t)_{\rm
  max}=2.0\cdot 10^{-4}$.  Consequently in this case the constraints
on $\Im Z_{ds}$ are only slightly changed relatively to scenario A,
and the upper bounds on the rare decays $\klpn$, $\kppn$ and $\kpe$
are not dramatically modified, as one can verify from tables
\ref{tab:eppgen} and \ref{tab:rareposgen}. On the contrary, for
negative $\Im \lambda_t$ and negative $\Im Z_{ds}$ there is the
possibility of a large cancellation in $\epp$ between the gluon and
electroweak penguin contributions leading to a possible strongly
enhanced value of $\Im Z_{ds}$. The upper bound on $-\Im Z_{ds}$ is
again given by \r{eq:imwds2} but now the most conservative bound is
obtained by setting $\Im\lambda_t=-2.0\cdot 10^{-4}$,
$B_8^{(3/2)}=0.6$, $B_6^{(1/2)}=1.2$ and varying $r_Z^{(8)}$ and $R_s$
in the ranges \r{eq:rz8} and \r{eq:rrs}. As shown in table
\ref{tab:eppgen} the maximal values of $-\Im Z_{ds}$ are this time
substantially larger than in scenario A and large enhancements of
$\klpn$ and $\kpe$ are possible as shown in table
\ref{tab:rareneggen}. Of course, this possibility seems quite remote,
since it would correspond to the case in which neutral meson mixing is
dominated by new physics contributions that are twice as large and
have opposite sign with respect to the Standard Model, and are
uncorrelated to the new physics contributions to $Z_{ds}$ (any
correlation between the different contributions can only tighten our
bound). However, if this situation could be realized in some exotic
model, then large enhancements of the branching ratios in question
would be possible. In figure \ref{fig:scenB} we plot the upper bound
on BR$(\klpn)$ for negative $\Im Z_{ds}$ as a function of $\Im
\lambda_t$, for three different values of $(\epp)^{\rm exp}_{\rm
  max}$. It is evident that as soon as one leaves the extreme case
$\Im \lambda_t = -2 \cdot 10^{-4}$ the enhancement of BR$(\klpn)$
quickly decreases.  

With $(\Re\lambda_t)_{\rm max}= -5.85 \cdot 10^{-4}$ the bound 
on ${\rm BR}(\kppn)$ in \r{bound} modifies to
\begin{equation}
\label{boundb}
{\rm BR}(\kppn)\le 0.229\cdot {\rm BR}(\klpn)+2.08\cdot 10^{-10}.
\end{equation}
This time the first term on the r.h.s is comparable to the second
term and a large enhancement of ${\rm BR}(\kppn)$ is possible. Simultaneously
the dependence on the experimental value of $\epp$ is stronger
than in the case of scenario B.

The most conservative bounds in scenario B are listed in
\r{scenb}.

\begin{table}
  \begin{center}
    \begin{tabular}{||c|c|c|c||}
      \hline \hline
      $(\epp)^{\rm exp}_{\rm max}$ & $2 \cdot 10^{-3}$ & $1.5 \cdot
      10^{-3}$ & $1 \cdot 10^{-3}$ \\ \hline 
      $\left(-\Im Z_{ds}\right)_{\rm max}[10^{-4}]$ 
      & $12.9~(6.8) $ & $11.0~(5.9)$ & $9.2~(4.9)$ \\ \hline \hline
      $(\epp)^{\rm exp}_{\rm min}$ & $0$ & $5 \cdot
      10^{-4}$ & $1 \cdot 10^{-3}$ \\ \hline
      $\left(\Im Z_{ds}\right)_{\rm max}[10^{-4}]$ 
      & $5.4~(3.1)$ & $4.5~(2.7)$ & $3.9~(2.4)$ \\ \hline \hline
    \end{tabular}
       \caption{Constraints on $\Im Z_{ds}$ from $\epp$ in scenario B
      for different
      values of $(\epp)^{\rm exp}_{\rm max}$ and $(\epp)^{\rm
        exp}_{\rm min}$, for $-2 \cdot 10^{-4} \le \lambda_t \le 2
        \cdot 10^{-4}$ with $B_8^{(3/2)}=0.6~(1.0).$} 
   \label{tab:eppgen}
  \end{center}
\end{table}

\begin{table}[htbp]
\begin{center}
  \begin{tabular}{||c|c|c|c||}
  \hline \hline
  $(\epp)^{\rm exp}_{\rm min}$ &  $0$ & $5 \cdot
  10^{-4}$ & $1 \cdot 10^{-3}$ \\ \hline 
  ${\rm BR}(\klpn)[10^{-10}]$ & $3.2~(1.4)$ &$2.4~(1.2)$ & 
   $2.0~(1.0)$\\ \hline 
  ${\rm BR}(\kpe)[10^{-11}]$ & $4.7~(2.1)$ &$ 3.4~(1.7)$ &
  $2.8~(1.5)$ \\ \hline 
  ${\rm BR}(\kppn)[10^{-10}]$ & $2.8~(2.4)$ & $2.6~(2.4)$ &
  $2.5~(2.3)$ \\ \hline \hline
  \end{tabular}
  \caption{Upper bounds for the rare decays $\klpn$, $\kppn$ and
    $\kpe$, obtained in scenario B by imposing the constraints in
    eq.~\r{boundb} and in table \ref{tab:eppgen}, in the case $\Im
    Z_{ds}>0$ with $B_8^{(3/2)}=0.6~(1.0).$}
  \label{tab:rareposgen}
\end{center}
\end{table}

\begin{table}[htbp]
  \begin{center}
    \begin{tabular}{||c|c|c|c||}
      \hline \hline
  $(\epp)^{\rm exp}_{\rm max}$ &  $2 \cdot 10^{-3}$ & $1.5 \cdot 10^{-3}$
  & $1 \cdot 10^{-3}$ \\ \hline 
  ${\rm BR}(\klpn)[10^{-10}]$ & $13.9~(4.6)$ & $10.5~(3.6)$&
  $7.6~(2.8)$ \\ \hline 
  ${\rm BR}(\kpe)[10^{-11}]$ & $21.9~(6.9)$ & $16.3~(5.4)$ &
  $11.7~(4.0)$ \\ \hline 
  ${\rm BR}(\kppn)[10^{-10}]$ & $5.3~(3.1)$ & $4.5~(2.9)$ &
  $3.8~(2.7)$ \\ \hline \hline
    \end{tabular}
    \caption{Upper bounds for the rare decays $\klpn$, $\kppn$ and
    $\kpe$, obtained in scenario B by imposing the constraints in
    eq.~\r{boundb} and in table \ref{tab:eppgen}, in the case $\Im
    Z_{ds}<0$ with $B_8^{(3/2)}=0.6~(1.0).$}
  \label{tab:rareneggen}
  \end{center}
\end{table}

\begin{figure}   
    \begin{center}
      \input{fig.tex}
    \end{center}
    \caption[]{Upper bounds on BR$(\klpn)$ in scenario B as a function
      of $\Im \lambda_t$, for negative $\Im Z_{ds}$ and for three
      different values of $(\epp)^{\rm exp}_{\rm max}$: $(\epp)^{\rm
      exp}_{\rm max} = 2 \cdot 10^{-3}$ (curve I), $(\epp)^{\rm
      exp}_{\rm max} = 1.5 \cdot 10^{-3}$ (curve II) and $(\epp)^{\rm
      exp}_{\rm max} = 1 \cdot 10^{-3}$ (curve III).}
    \label{fig:scenB}
\end{figure}

\subsection{Constraints from $\epp$ in scenario C}
\label{sec:caseC}
Finally we consider scenario C, which corresponds to scenario B
 in which the CKM matrix is real
and $\Im\lambda_t$ vanishes. CP-violation originates then 
entirely in new physics contributions. In this case we
find

\begin{equation}
{\rm BR}(\klpn)=6.78\cdot 10^{-4} \left[\Im Z_{ds}\right]^2
\end{equation}
\begin{equation}
{\rm BR}(\kpe)=1.20\cdot 10^{-4} \left[\Im Z_{ds}\right]^2
\end{equation}
and the bound \r{boundb}. If $(\epp)^{\rm exp}$ is positive then
only $\Im Z_{ds}<0$ is allowed. Then we have
\begin{equation}
  \label{eq:mwds5}
  -\Im Z_{ds} \le \frac{\left(\frac{\varepsilon^\prime}{\varepsilon}
  \right)^{\rm exp}_{\rm max}}
 {\left(6.5~ B_8^{(3/2)} -   1.20\right) } 
\end{equation}
and the numerical results in table \ref{tab:rarec}. This time the
upper bounds on ${\rm BR}(\klpn)$ and ${\rm BR}(\kpe)$ are
substantially stronger than in the case of scenario B.  The most
conservative bounds in scenario C are listed in \r{scenc}.

\begin{table}[htbp]
  \begin{center}
    \begin{tabular}{||c|c|c|c||}
      \hline \hline
  $(\epp)^{\rm exp}_{\rm max}$ &  $2 \cdot 10^{-3}$ & $1.5 \cdot 10^{-3}$
  & $1 \cdot 10^{-3}$ \\ \hline 
$\left(-\Im Z_{ds}\right)_{\rm max}[10^{-4}]$ 
& $7.4~(3.8)$ & $5.6~(2.8)$&
  $3.7~(1.9)$ \\ \hline
${\rm BR}(\klpn)[10^{-10}]$ & $3.7~(1.0)$ & $2.1~(0.5)$&
  $0.9~(0.2)$ \\ \hline 
  ${\rm BR}(\kpe)[10^{-11}]$ & $6.6~(1.7)$ & $3.7~(1.0)$ &
  $1.6~(0.4)$ \\ \hline 
  ${\rm BR}(\kppn)[10^{-10}]$ & $2.9~(2.3)$ & $2.6~(2.2)$ &
  $2.3~(2.1)$ \\ \hline \hline
    \end{tabular}
    \caption{Upper bounds for the rare decays $\klpn$, $\kppn$ and
    $\kpe$, obtained in scenario C
    in the case $\Im Z_{ds}<0$ with $B_8^{(3/2)}=0.6~(1.0).$}
  \label{tab:rarec}
  \end{center}
 \end{table}

\section{Summary}
\label  {sec:summary}

In this paper we have considered the possibility of an enhanced $\bar
s d Z$ vertex $Z_{ds}$. We have pointed out that in spite of large
theoretical uncertainties the best present constraints on $\Im Z_{ds}$
follow from $\epp$. These constraints should be considerably improved
when new data on $\epp$ will be available and the theoretical
uncertainties reduced. While considerable enhancements of ${\rm
  BR}(\klpn)$ and ${\rm BR}(\kpe)$ over the Standard Model
expectations are still possible, the huge enhancements by two orders
of magnitude claimed in \cite{CI} are in our opinion already excluded.
This is in particular the case in scenarios in which the analysis of
the unitarity triangle is only insignificantly modified by new physics
contributions.

Similar comments apply to ${\rm BR}(\kppn)$ which is dominantly
bounded by ${\rm BR}(\kmm)_{SD}$. Using the most recent estimates of
the long distance dispersive contribution to ${\rm BR}(\kmm)$
\cite{dambrosio,pich} we find that values of ${\rm BR}(\kppn)$ of the
order of $10^{-9}$ are certainly excluded.  In this
context we have presented an analytic upper bound on ${\rm BR}(\kppn)$
as a function of ${\rm BR}(\klpn)$ and ${\rm BR}(\kmm)_{SD}$.

Our results for various bounds are collected in tables 1-7 and figure
1 with the most conservative bounds listed in \r{eq:bounded},
\r{eq:boundedp}, \r{scenb} and \r{scenc}.  Clearly the best
constraints on $\Im Z_{ds}$ and $\Re Z_{ds}$ will follow in the future
from precise measurements of the theoretically clean branching ratios
${\rm BR}(\klpn)$ and ${\rm BR}(\kppn)$ respectively. Meanwhile it
will be exciting to follow the development in the improved
experimental values for $\epp$, ${\rm BR}(\kmm)$, ${\rm BR}(\kppn)$,
${\rm BR}(\klpn)$ and ${\rm BR}(\kpe)$.

\section*{Acknowledgements}

We warmly thank Gilberto Colangelo, Gino Isidori and Yossi Nir for
discussions and clarifications abour their work.  We would also like
to thank Gustavo Burdman, Louis Fayard, Laurie Littenberg and Bruce
Winstein for asking questions related to \cite{CI}, which motivated
this work.

\end{document}